\documentclass[%
 reprint,
 amsmath,amssymb,
 aps,
]{revtex4-2}

\usepackage{hyperref}% add hypertext capabilities
\usepackage{orcidlink}
\usepackage{comment}
\usepackage{graphicx}% Include figure files
\usepackage{dcolumn}% Align table columns on decimal point
\usepackage{bm}% bold math
\usepackage{booktabs}
\usepackage{xcolor}
\usepackage{soul}
\begin{document}

\preprint{APS/123-QED}

\title{Simulations of Classical Three-Body Thermalization in One Dimension}% Force line breaks with \\

\author{M. Eltohfa}
 \email{meltohfa@purdue.edu}
 \affiliation{
 Department of Physics and Astronomy, Purdue University, West Lafayette, Indiana 47906 USA
}
%^{\orcidlink{<https://orcid.org/0000-0000-0000-0000>}}

%\item \href{https://orcid.org/0000-0000-0000-0000}{\textcolor{orcidlogocol}{\aiOrcid} \hspace{2mm} orcid.org/0000-0000-0000-0000}

% ORCID https://orcid.org/0009-0000-6296-5412

\author{Xinghan Wang}%
 %\email{wang3813@purdue.edu}
\affiliation{
 Department of Physics and Astronomy, Purdue University, West Lafayette, Indiana 47906 USA
}
% Xinghan Wang
%orcid: https://orcid.org/0009-0006-0313-7952

\author{Colton M. Griffin}%
% \email{griff254@purdue.edu}
\affiliation{
 Department of Physics and Astronomy, Purdue University, West Lafayette, Indiana 47906 USA
}
%preferred name: Colton M. Griffin
%Colton Griffin name for Orcid ID
% Colton M. Griffin
%orcid: https://orcid.org/0000-0003-2558-0219

\author{F. Robicheaux}%
 \email{robichf@purdue.edu}
\affiliation{
 Department of Physics and Astronomy, Purdue University, West Lafayette, Indiana 47906 USA
}
%orcid: https://orcid.org/0000-0002-8054-6040

\begin{comment}
\collaboration{MUSO Collaboration}%\noaffiliation

\author{Charlie Author}
 \homepage{http://www.Second.institution.edu/~Charlie.Author}
\affiliation{
 Second institution and/or address\\
 This line break forced% with \\
}%
\affiliation{
 Third institution, the second for Charlie Author
}%
\author{Delta Author}
\affiliation{%
 Authors' institution and/or address\\
 This line break forced with \textbackslash\textbackslash
}%

\collaboration{CLEO Collaboration}%\noaffiliation
\end{comment}

\date{\today}% It is always \today, today,

\begin{abstract}

One-dimensional systems, such as nanowires or electrons moving along strong magnetic field lines, have peculiar thermalization physics. The binary collision of point-like particles, typically the dominant process for reaching thermal equilibrium in higher dimensional systems, cannot thermalize a 1D system. We study how dilute classical 1D gases thermalize through three-body collisions. We consider a system of identical classical point particles with pairwise repulsive inverse power-law potential $V_{ij} \propto 1/|x_i-x_j|^n$ or the pairwise Lennard-Jones potential. Using Monte Carlo methods, we compute a collision kernel and use it in the Boltzmann equation to evolve a perturbed thermal state with temperature $T$ toward equilibrium. We explain the shape of the kernel and its dependence on the system parameters. Additionally, we implement molecular dynamics simulations of a many-body gas and show agreement with the Boltzmann evolution in the low density limit. For the inverse power-law potential, the rate of thermalization is proportional to $\rho^2 T^{\frac{1}{2}-\frac{1}{n}}$  where $\rho$ is the number density. The corresponding proportionality constant decreases with increasing $n$.

\begin{comment}
\begin{description}
\item[Usage]
Secondary publications and information retrieval purposes.
\item[Structure]
You may use the \texttt{description} environment to structure your abstract;
use the optional argument of the \verb+\item+ command to give the category of each item. 
\end{description}
\end{comment}

\end{abstract}

\maketitle

%\tableofcontents

\section{Introduction}

\subsection{Technological Motivation}
With the advance of technology, very thin systems can be produced that can be approximated as one-dimensional (1D). Examples of 1D gases include quantum-wires made of GaAs \cite{christen1992ultrafast} or carbon nanotubes \cite{CNTs}. Other examples come from plasma physics where electrons are confined to move along strong magnetic field lines \cite{o1985collisional}, or isolated, far-from-equilibrium, Bose gases \cite{erne2018universal}. Thus, the properties of 1D systems are of some interest.

One dimensional confinement considerably affects system properties, such as thermalization process \cite{movsko1994thermalization}, enhanced correlations and collective behavior \cite{haldane1981luttinger}, and anomalous transport and diffusion \cite{livi2022anomalous}. In the present work, we study the rate of thermalization of certain 1D systems. This rate measures how fast the equilibrium state is reached if the system starts from a non-equilibrium state. From another perspective, it measures how fast the system loses memory of its initial state.
%Technologically, systems with low thermalization rates are quite useful, for example, to construct high-fidelity memory elements for computers.

Unique features of 1D thermalization have been demonstrated. Optical measurements have shown that carrier relaxation is much slower in quantum wires than in bulk and two-dimensional forms \cite{rota1993reduced}. Additionally, molecular dynamical (MD) simulations of a 1D line of electrons have shown the system thermalizes in the order of $10\, ns$ \cite{movsko1994thermalization}, which is a relatively slow rate. On the other hand, the dynamics of thermalization of quasi-1D systems, consisting of nearly decoupled chains, was shown to exhibit non-exponential approach to equilibrium \cite{panfil2023thermalization}. 

For this paper, we simulate the interaction of classical point-like particles that are confined to move in 1D.  We define the thermalization rate by how fast a special velocity of a particle irreversibly diffuses into the distribution of the rest of the system. With this definition, we show that the 1D gases under study exhibit very slow thermalization rates.

 %definition thermalization rate
 %memory of intitial condition with great fidelity is ralated to rate of thermalization
 %systems that resist thermalization https://en.wikipedia.org/wiki/Thermalisation

\subsection{Theoretical Motivation}

\subsubsection{Two-body thermalization}
 In the kinetic theory of 3D gases, thermalization through binary collisions has been thoroughly studied \cite{uhlenbeck1963lectures}. Binary collisions are the dominant interaction if the gas is dilute; many-particle collisions are suppressed by powers of the density \cite{Landau}. Although a binary collision is tightly constrained by several conservation conditions, there is freedom for the particles to change their directions based on their impact parameter. 
 
 In the center of mass frame of the two colliding bodies, the equations of motion can be integrated and a non-zero differential cross-section can be obtained. Subsequently, the cross section determines the rate of scattering from and into a tiny volume in coordinate-velocity space of the one-particle distribution. These rates can be used in the Boltzmann Equation to propagate the distribution \cite{uhlenbeck1963lectures}. By Boltzmann's $H$ theorem, a non-zero cross section guarantees the thermalization of the system to equilibrium. Binary collisions also lead to thermalization in 2D although the details are different.

%The problem is to be treated classically assuming that the thermal de Broglie wavelength is small compared to the inter-particle distance. That is the case for hot gases. The Plasma factor and correlations, The dynamical temperature Q in paper
Thermalization through binary collisions, however, does not work in a 1D gas of identical point like particles. Consider two particles of mass $m$ elastically colliding with incoming velocities $v_1$, $v_2$, and outgoing velocities $w_1$ and $w_2$. Conservation of energy and momentum must hold. This, in 1D, entails two equations which completely determine that $w_2=v_1$ and  $w_1=v_2$. This is a trivial swapping that leads to the same velocities and, therefore, to an unchanged velocity distribution\cite{movsko1994thermalization}.

The triviality of the binary collision in 1D hinges on two assumptions \cite{movsko1994thermalization}: 1- the dispersion relation is parabolic which is well established. 2- there is no exchange of momentum with the substrate (the crystal or the medium the particles live in). The latter assumption works well in regimes of energy less than $1 eV$ for crystal spacing $10^{-10}m$, or if the particles live in 1D vacuum. In this work, we assume that the particles are constrained to move on a ring or a line with no external forces.

\subsubsection{Many-body thermalization in 1D}
It was shown in \cite{movsko1994thermalization} that many-body Coulomb scattering can thermalize a one-dimensional electron gas in a single-subband GaAs quantum wires. This was done through classical molecular dynamics (MD) simulations. In the study, the gas is dense enough such that the mean potential energy is of the order of the mean kinetic energy $\sim 100K \sim 10meV$. It was calculated that the relaxation time is of order $10\,ns$ and increases rapidly for lower densities. 
%$ke^2/\lambda$, where $\lambda$ is the mean inter-particle distance,

\subsubsection{Three-body thermalization in 1D}
 %Many-body thermalization, however, is computationally expensive to study because of the large number of particles involved. It is also hard to study analytically because of the lack of conserved quantities and the inseparability of the various degrees of freedom.
Since binary collisions cannot thermalize a 1D system, we study thermalization through the next simplest process, the three-body (ternary) collisions. In very dilute gases, which is our main focus, the ternary collision is dominant over the higher order collisions. The ternary collision is generally non-trivial and can generate new velocity states.  For long-range interacting homogeneous 1D systems, it was shown that they can thermalize through 3-body effects, but their relaxation is drastically slowed down \cite{fouvry2020kinetic}. The 3-body problem, however, is non-integrable \cite{maciejewski2011non} (except in very special cases \cite{calogero1974exact}), so we study its trajectories numerically.

%The recent advances in of plasma physics where electrons move along a magnetic field line and technological advances in nano tubes \cite{} have made 1 dimensional gas models more popular. Thus it is necessary to study their thermalization mechanism through the 3 body problem.

Three-body thermalization has also been addressed in \cite{ma1983one} for a model problem. The scattering rate from a triple of initial momenta to a triple final momenta was assumed for simplicity to be constant as long as the incoming momenta and the outgoing momenta satisfy energy and momentum conservation. In such cases, the Boltzmann Eq. (\ref{eqn:Boltz}) is exactly solvable. This constant scattering rate, however, was not derived from an inter-particle potential energy.  %Therefore, it is not clear which physical systems the model is applicable to.
Additionally, it was found that the rate of collisions, and hence the rate of thermalization, goes as $\rho^2$ but is not affected by temperature or the average kinetic energy. With the rates computed from an inter-particle potential, we will show that if we start from a quasi-thermal distribution of temperature $T$, the rate of thermalization depends not only on $\rho$ but also on $T$.

\subsection{Goal and Plan}

%in the inverse power potential.

In the present paper, we first introduce a model of a 1D gas on a ring and a model of thermalization.  We consider a system with pairwise inverse power potential with power $n \geq 2$. This potential is formally long-range, since any particle can affect any other particle with non-zero force. But for low densities and high temperatures, which we assume, this force is small enough that long-range effects can largely be ignored \cite{fouvry2019kinetic}. We consider the evolution of a `delta-perturbed' thermal state. Previous studies \cite{movsko1994thermalization}, \cite{panfil2023thermalization} considered the evolution of a bimodal distribution or a modified Gaussian \cite{fouvry2019kinetic}.  

Second, we implement MD simulations and discuss the scaling of the thermalization rate, which we define as the initial rate of the spread of the perturbation. Third, using Monte Carlo simulations, we calculate a three-body collision kernel for the inverse power potential and the Lennard-Jones potential. Using the kernel information, we show how the transition rates scale with $\rho$ and $T$ and compute the thermalization rate for a range of parameters. Fourth, the collision kernel is used in the Boltzmann equation to evolve the perturbation and this method is shown to be in agreement with the MD simulations. Finally, we discuss the shape of the kernel and how it changes with the inverse potential power $n$. 
\section{Gas and Thermalization Model } \label{section:modelling the gas}

\begin{figure}[h]
\resizebox{75mm}{!}{\includegraphics{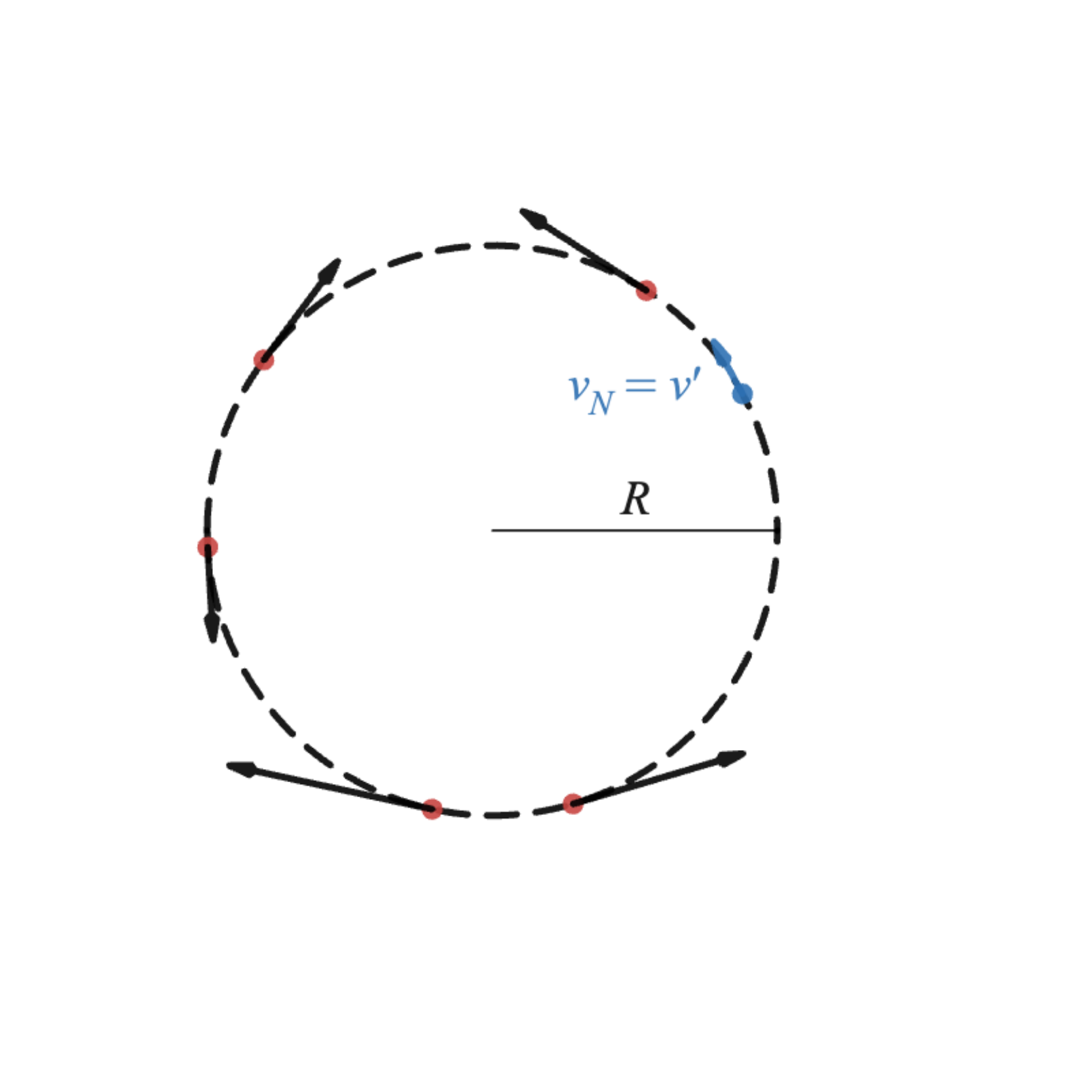}}
\caption{\label{fig:ring} Particles on a ring of radius $R$. All but one particle are initialized from a thermal distribution. The `special' particle (in blue) is initialized with a special velocity $v'$. The special particle represents a perturbation to the thermal state.}
\end{figure}

We consider $N$ identical particles of mass $m$ constrained to move on a ring of radius $R$ as in Fig. \ref{fig:ring} with pairwise repulsive inverse power-law potential
\begin{equation}
\label{eqn:ipp}
    U(d)=\epsilon\left(\frac{l_0}{d}\right)^n,
\end{equation} where $d$ is the pairwise separation   {(the 2D distance in the plane of the ring)}, $l_0$ is the basic unit of length at which the potential energy is equal to some interaction strength $\epsilon > 0$, and $n$ is an even integer $\geq2$.

% Refree 1, Edit 1%
The locations of the particles are parameterized by the angles $\phi_i$ and the 2D distance   {(the length of the chord) between particles $i$ and $j$ is} 
\begin{equation}
   d_{ij}= 2\,R\,\sin\left(\frac{|\phi_i-\phi_j|}{2}\right).
   \label{eq:distance}
\end{equation}
The energy of the system is
\begin{equation} \label{eqn:Hamiltonian}
    E = \sum_{i=1}^{N} \frac{mR^2\omega_i^2}{2} + \epsilon\left(\frac{l_0}{2R}\right)^n \sum_{i,j>i}^{N}  \left[\sin\left(\frac{\phi_i-\phi_j}{2}\right)\right]^{-n},
\end{equation} where $\omega_i$ is the angular velocity of particle $i$.

We choose a ring to represent the 1D system instead of a line so that particles do not escape to infinity under repulsive forces. Modeling a line would require a confining potential which leads to the particles at the edges experiencing a different environment from those in the center of the range.   {Although the aim of the paper is studying the gas in 1D, we use the 2D distance along the chord} in Eq. (\ref{eq:distance})   {instead of the 1D distance along the arc so that the torque would be continuous in $\phi$ and would be $0$ for particles on opposite sides of the ring. It also has the added benefit of smoothly cutting off the interaction between two particles as $|\phi_i - \phi_j|$ increases. This introduces curvature effects which distort from a true 1D system. However, these effects become negligible at large $R$. The choice of a finite ring with periodic boundary also introduces finite size effects that diminish at large $R$ and $N$. To simulate a particular linear number density, $\rho = N/R$, we decrease the ring effects by doubling $R$ and $N$ until our results converge.}

%We are aware of the two artifacts of the ring: curvature and periodic effects. To simulate a particular linear number density, $\rho$, we tackle the ring effects by doubling $R$ and the number of thermal particles, $N-1$, until our results converge.$

We consider a thermal initial condition sampled from a Boltzmann distribution with temperature $T$.  
% the potential is week, so we can consider non-interacting particles and leads to maxwell-boltzmann statistics?
In particular, the initial velocity $v(t=0)=R \, \omega (t=0)$ of a particle follows the Maxwell-Boltzmann statistics with probability density function:
\begin{equation} \label{eqn:MB distribution}
    f_0(v) =\frac{1}{\sqrt{2\pi v_{th}^2}}e^{-v^2/2v_{th}^2}
\end{equation}
\begin{comment}
    \begin{equation}
    v\sim \mathcal{N}(0,\,v_{th}^{2}),
\end{equation}
where $\mathcal{N}$ is the normal distribution and
\end{comment}
 where $v_{th} = \sqrt{kT/m}$ is the velocity scale set by the temperature.  $k$ is the Boltzmann constant.
% Refree 2 second Question
 {  {The initial locations are chosen randomly according to a relative potential energy Boltzmann factor $exp(-PE/kT)$, where $PE$ is the total potential energy} (second term in Eq. (\ref{eqn:Hamiltonian})).   {In this way, a configuration with a very close group of particles occurs rarely and, therefore, rejected with a high probability in our simulations. Also, the number density $\rho$ is homogeneous on average. In our MD simulations, we ensure uniform density by averaging the results over many runs with different initial conditions. We do not consider effects of density variations along the ring (which lead to the diffusion of the particles) in this paper. However, we will show how the relaxation rate depends on the overall density.}}
 
% Refree 1 Edit 2
In this work, we assume high enough $T$ and small enough number density, $\rho$, so that the average kinetic energy is much greater than the potential energy (unlike in Ref. \cite{movsko1994thermalization}, where the energies are comparable).   {In this regime, correlations between positions of the particles can be ignored. At the other extreme of low $T$ and high $\rho$, position correlations become important because the system condenses to a solid with periodic arrangement of particles.}

%, higher order collisions, and collective effects can be ignored as in Ref. \cite{fouvry2019kinetic}.

%\subsection{Thermalization Model}
%\label{sec:thermalization model}

We consider a perturbation to the thermal state by forcing   {one of the particles (e.g. the $N^{th}$ particle)} to start at a particular velocity $v'$ as in Fig. \ref{fig:ring}. While all the rest of the phase variables obey the thermal distribution statistics, this particle's velocity will disobey the Maxwell Boltzmann distribution in Eq. (\ref{eqn:MB distribution}) and will rather have a delta function distribution centered at $v'$. We refer to such a particle as the `special' particle and its velocity as the `special velocity', while we refer to the rest of the particles as `thermal'. The velocity probability density function of the total system is

\begin{eqnarray} \label{eqn:initial distribution}
    f(v,t=0)
    &=& \frac{N-1}{N}f_0(v)+\frac{1}{N}h(v,t=0),
\end{eqnarray} where $h(v,t)$ is the normalized perturbation (integral over $v$ is $1$) such that
\begin{equation}
    h(v,t=0) = \delta(v-v').
\end{equation}

If we introduce the scaled velocity $u= v/v_{th}$, then the normalized (integral over $u$ is $1$) initial distribution is
\begin{eqnarray} \label{eqn:normalized intitial}
    f(u,t=0)
    &=& \frac{N-1}{N}\frac{1}{\sqrt{2\pi }}e^{\frac{-u^2}{2}} + \frac{1}{N} \delta(u-u'),
\end{eqnarray}

Our goal is to study the evolution of $h$. According to the Boltzmann $H$-theorem \cite{uhlenbeck1963lectures}, the steady state distribution $f(v,\infty)$, and therefore $h(v,\infty)$, is the equilibrium distribution $f_0$. This is strictly true in the thermodynamic limit $N\rightarrow \infty$. For finite $N$, however, the steady state has a slightly different temperature $T_f\simeq (N-1+u'^2)T/N \approx T$ for large $N$, which is a result of energy conservation.
% Refree 1 Edit 3

%Introduce the normalization to u here.

%On what time scale does the energy of the perturbation grows?

%According to Boltzmann's H theorem, the  evolves to become a Gaussian. From our simulations we show the spike evolving like in figure %include figure

%What is $\Gamma$ and how it scales with the system parameters?
%What is the proportionality constant?
%include time series like professor sugguested. 

%Figure \ref{fig:MD_with_time} shows a time series of the evolved perturbation (delta function at initial time) for  $N=41$,  $R=6400\,\mu m, t_f=3200\,ns$. 
%Figure \ref{fig:MD_with_time} shows the evolution of the delta perturbation of a system with parameters 

To investigate the rate of thermalization, we introduce $\Gamma$ to be the initial rate of change of the variance of the perturbation:
\begin{equation} \label{eqn: Gamma definition}
    \left.\Gamma = \frac{d\langle{(u-u')}^2\rangle_h}{dt} \right\vert_{t=0}.
\end{equation}
This definition of  $\Gamma$  implies that if the perturbation continues spreading at the initial rate, then it takes time of order $1/\Gamma$ for the perturbation $h(u)$ to reach variance $\sim 1$, i.e., the special particle reaching the temperature of the bath. This definition is introduced to enable the quantification of the notion of the thermalization rate.

  {We choose this particular type of perturbation because it is easy to track and characterize during the simulations. The initial perturbed distribution is a Gaussian with a large narrow spike at $v'$. When the number of particles is large, the Gaussian background is mostly unaffected during the evolution. On the other hand, during the early stages of the thermalization the spike spreads out but remains significantly higher than the background Gaussian. It is therefore possible, during early times, to track the evolved perturbation by only tracking a small range of velocities (we choose a range of $ 0.2 v_{th} $ centered around $v'$) and subtracting the background to get $h$. This eliminates the need to make a histogram of all the velocities and makes it possible to get a clearer evolution of the perturbation. Although our choice is mostly for ease of computations, it will be shown below that the relaxation rate, $\Gamma$, is only different by a small numerical factor for different types perturbations.}
%Refree 1 Edit 4
% 4. The caption of Figure 2 could be written more clearly: I had to re-read it a couple of times just to understand what it meant. It is awkward that the caption refers to \Gamma before this quantity has been defined in the text, and it is confusing that h(v,t) is only defined as "the perturbation"—a more specific description as the probability distribution of the velocity of the Nth particle would be clearer. I would also appreciate some discussion of why the blue and green curves in Figure 2 appear so noisy. I normally expect probability distributions to be quite smooth; samplings of the distributions will yield a lot of different specific values, but for a probability distribution itself to be so noisy seems to me to be almost a contradiction in terms. Are these very noisy probability distributions somehow inferred from a set of samples? If so, perhaps some better inference procedure should be used, which will give smoother inferred probability distributions.

%need inference rules to get a smoother probablity curve

\begin{figure}[h]
\resizebox{86mm}{!}{\includegraphics{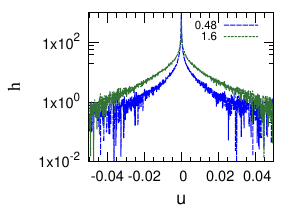}}
\caption{\label{fig:MD_with_time}   {MD simulated evolution of the probability distribution of the perturbation $h(v,t)$ which is initially localized at $v = 0$. The two times shown are $0.48$ and $ 1.6$ $ (\times 10^{-4}/\Gamma$) which are very early in the thermalization process. The $x$-axis is the scaled velocity $u = v/v_{th}$. Comparing the distribution at these two time instants and at $t=0$ (a delta function of $u$), the perturbation width increases with time as its height drops. The fluctuations at the edges of the curves are due to statistical noise.}}
\end{figure}

% Refree 1 Edit 4
% notice on the natural scale fig 2 would look like fig 3. Noise is so much amplified on the log scale.
We run many MD simulations (as described in detail in Appendix \ref{sec:MD simulation}) where the $N-1$ thermal particles are randomly chosen from a thermal distribution. Averaging over these many runs gives an average $h(u,t)$. From our simulations, the average $h(u,t)$ starting from $\delta (u)$ is shown for two times in Fig. \ref{fig:MD_with_time}. As expected, the perturbation's peak height decreases and its width increases as the system evolves. The times shown are early in the thermalization process where
the population only spreads to small velocities ($|u|< 0.06$). However, the cusp feature of the delta function is maintained throughout this time period, which is an indication that the gas is far from equilibrium. The parameters of the simulated gas are $l_0=1\,\mu m$, $\epsilon= 1\,eV$, $kT= 100\,meV$, $m=m_e$ (mass of the electron), $n=6$, $N=41$, and $R=6400\,\mu m$. The times shown are $t_f= 960, 3200\,ns = 0.48, 1.6 \times 10^{-4}/\Gamma$, where $\Gamma$ is computed in Eq. (\ref{eqn: rate initial}).
%($\Gamma$ being the rate of thermalization defined below in Eq. (\ref{eqn: Gamma definition}))

 In the rest of the paper, we show how $\Gamma$ depends on the system parameters $\{R, N, m, T, l_0, \epsilon, n, u'\}$. To extract this dependence, we do MD simulations for different parameter sets; however, as argued in Appendix \ref{sec:MD simulation}, such $N$-body simulations are computationally very expensive when $N$ is large or the propagation time, $t_f$, is large.

%We are interested the scaling of $\Gamma$ with the system's parameters  and calculate the proportionality constant. In addition, we study the behavior with the potential power $n$.

%\subsection{Results of MD Simulations}

\section{Boltzmann Evolution of the Gas}
\label{Markov}

 A more efficient method to study $\Gamma$ is to consider the simpler process by which our system thermalizes, the ternary collision. By knowing the frequency of such collisions and how they change the particle velocities, we can propagate the velocity distribution in time. The recipe of such method is the Boltzmann equation. For those reasons, we choose to focus on the Boltzmann method and limit ourselves to a few MD simulations. In particular, we use the MD simulations to extract some preliminary scaling for $\Gamma$ and as a benchmark to verify our Boltzmann calculations.

The Boltzmann evolution is an example of a continuous time Markov chain in which the next step distribution is only dependent on the current distribution.  If the perturbation is initially localized at $v'$, $h(v,0)=\delta(v-v')$, then after infinitesimal duration $dt$, the perturbation becomes
\begin{equation} \label{eqn:CTMC_localized}
     h(v,dt) = \delta(v-v')+ dt  K_{v'\rightarrow v},
\end{equation} where the kernel $K_{v'\rightarrow v}$ is the rate of transitioning from velocity $v'$ to the range between $v-\frac{\Delta v}{2}$ and $v+\frac{\Delta v}{2}$ per $\Delta v$ in the limit $\Delta v \to 0$ . After a finite time, the perturbation delocalizes to a continuous range of $v$; the evolution equation then becomes

\begin{comment}
    
\begin{equation}
     f(v,dt)\Delta v = f(v,0)\Delta v + dt  \int_{-\infty}^{\infty} \Delta v G_{v'\rightarrow v} f(v',0) \,dv' 
\end{equation}
\end{comment}
\begin{equation} \label{eqn:CTMC}
     h(v,t+dt) = h(v,t)+ dt  \int_{-\infty}^{\infty} K_{v'\rightarrow v} h(v',t) \,dv' 
\end{equation}
%     D(v,dt)\Delta v = D(v,0)\Delta v + dt \[ \int_{-\infty}^{\infty} \Delta v G_{v'\rightarrow v} D(v',0) \,dv' \]
%The integral equation is difficult to solve numerically
 Numerically, we work with a discretized version of the distribution and the kernel. The velocity axis is divided into bins of width $\Delta v$, which we choose as a fraction of $v_{th}$. Each bin is labeled by its center velocity $v_i = i \Delta v$ and extends from $v_i-\frac{\Delta v}{2}$ to $v_i+\frac{\Delta v}{2}$. Equation (\ref{eqn:CTMC}) becomes
% Refree 1 Edit 5
\begin{equation} \label{eqn:Dis_CTMC}
     h_i(t+dt) \approx h_i(t) + dt \sum_{j=-\infty, j\neq i}^{j=\infty} \Delta v \left[K_{v_j\rightarrow v_i} h_j(t) - K_{v_i\rightarrow v_j} h_i(t)\right],
\end{equation}where $h_i \Delta v$ is the population in bin $v_i$. The positive term in the square brackets represents the flow of population to bin $v_i$ from all other bins, while the negative term represents the flow out of bin $v_i$ to all other bins.

\subsection{Monte Carlo Simulation of $K$}
\label{sec:MC}

To calculate the transition rates $K_{v'\rightarrow v}$, we focus on the special particle with velocity $v'$ and treat the rest of the system as a thermal bath of temperature $T$ and number density $\rho$. Through a Monte Carlo (MC) simulation on a line \cite{kroese2013handbook}, we simulate the possible collisions the special particle (referred to as particle 1) encounters with two other thermal particles (referred to as particles 2 and 3).
 %to calculate the transition rates from the special velocity $v'$ to other velocities $v$. 

%We distinguish between 3 cases:
%frozen perturbation (frozen special particle)
%cold perturbation
%hot perturbation
%Single trajectory dynamics

The set of all possible 3-body collisions can be determined by first changing to an inertial frame moving with velocity $v'$ where the special particle is at rest before the collision. In this frame, particle 1 is initialized with velocity $v_1=0$ at the origin $x=0$. We consider an observation region of length $L$ centered around the special particle. In a small time duration, there is a probability that a thermal particle (particle 2 or 3) will enter the observation region from either side at $x=\pm L/2$ with some velocity ($v_2$ or $v_3$). We model the `launching' of the particles into the observation region as a Poisson process with rate $r$. The rate $r$ determines the distribution for the delay $\Delta t$ between the launched particles.  By studying the phase space distribution of a particle in the considered thermal bath, we determine $r$ and the statistics of the launched particles. These statistics and the steps of the algorithm are described in Appendix \ref{Appendix:MC Algorithm}.

   {This idea of tracking a particle in contact with a bath in thermal equilibrium is similar to the analysis of Brownian motion }\cite{Zwanzig_2009}.   {But instead of following the position of a tagged particle to see how it diffuses in space (like in a random walk), we track the special velocity and see how it changes due to 3-body collisions. By studying the velocity fluctuations and their rates, we can infer the relaxation rates of macroscopic perturbations. This is similar to using the time correlation function of a particle velocity in Brownian motion to calculate the diffusion coefficient}\cite{Zwanzig_2009}.   {Deducing the thermalization rate from the velocity changes is an application to the Onsager regression hypothesis which relates macroscopic non-equilibrium disturbances to microscopic fluctuations in the corresponding equilibrium system}\cite{j2007statistical}.

\section{Results}
\subsection{Scaling Behavior of the $N$-body Gas}
 
In this section, we extract the 3-body scaling using MD simulations. Before presenting the results of the simulations, we predict the scaling by analyzing the equations of motion of the $N$-body gas on the ring.
The equations of motion follow from the energy in Eq. (\ref{eqn:Hamiltonian}): 
\begin{eqnarray} \label{eqn:dynamical equations ring}
    mR\frac{d\omega_i}{dt} 
&=&\sum_{j\neq i}^{N} \frac{n\epsilon}{2}\frac{\left(\frac{l_0}{2R}\right)^n}{R\left[\sin\left(\frac{\phi_i-\phi_j}{2}\right)\right]^{n+1}}\cos\left(\frac{\phi_i-\phi_j}{2}\right), \nonumber\\
    \frac{d\phi_i}{dt} &=& \omega_i.
\end{eqnarray} These equations can be scaled resulting in equations of motion independent of all dimensional parameters.

We reduce the number of parameters by first identifying the length, time, and angular velocity scales. The angular velocity scale comes naturally from the thermal velocity: $\omega_{th}=v_{th}/R$, where $v_{th}$ was defined in the context of Eq. (\ref{eqn:MB distribution}). The dynamical time scale is proportional to the average orbital time: $t_d = 1/\omega_{th}$. The simple form of the inverse power potential is utilized to find the length scale. The pairwise interaction can be rewritten as
\begin{equation} \label{eqn:inverse power potential-kT scaled}
    U(d)=kT\left(\frac{l}{d}\right)^n,
\end{equation} where
\begin{equation} \label{eqn:length scale}
    l=l_0\left(\frac{\epsilon}{kT}\right)^\frac{1}{n}
\end{equation} is the length scale (closest approach distance) set by the temperature. 
%In other words, the potential itself is scale free, and the scale is set by the kinetic energy.

If the scaled angular velocity is $\tilde{\omega}=\omega/\omega_{th}$ and the scaled time is $\tilde{t}=t/t_d$ then 
\begin{eqnarray} \label{eqn:scaled_ring_dynamics}
    \frac{d\tilde{\omega}_i}{d\tilde{t}} &=& \sum_{j\neq i}^{N} \frac{n}{2}\frac{c}{\left[\sin\left(\frac{\phi_i-\phi_j}{2}\right)\right]^{n+1}}\cos\left(\frac{\phi_i-\phi_j}{2}\right), \nonumber\\
    \frac{d\phi_i}{d\tilde{t}} &=& \tilde{\omega}_i,
\end{eqnarray} 
where 
\begin{equation}
c = \left(\frac{l}{2R}\right)^n.
\end{equation}
The dynamics of two systems in terms of the scaled variables is identical if their corresponding $c$, $n$, and $N$ parameters are the same. The disappearance of the energy scales ($kT$ or $\epsilon$) in the scaled Eq. (\ref{eqn:scaled_ring_dynamics}) is owed to the scale invariance of the inverse power potential. In Sec. \ref{sec: LJ}, we demonstrate how the energies re-enter the dynamics if we consider other potentials such as the Lennard-Jones potential, which does not lead to scaled equations of motion.

% Refree 1 Edit 6
From the definition of $l$ in Eq. (\ref{eqn:length scale}), $kT$ scales as $l^{-n}$. Furthermore, the density $\rho$ of the particles for a given $N$ scales as $R^{-1}$. Thus, $c$ scales as $T^{-1}\rho^n$. If one scales ${\rho} {\rightarrow}a\rho$ and $T{\rightarrow}a^nT$,  $c$ remains the same. Therefore, the scaled dynamics (Eq. (\ref{eqn:scaled_ring_dynamics}) alongside the scaled initial conditions) remains the same .   {Given this constraint, one can show that any time unit of the form $\rho^{-s} T^{-(\frac{1}{2}-\frac{s-1}{n})}$ would be invariant under our scaling symmetry. An invariant rate unit $\gamma$ is thus governed by terms of the form (rewritten using $l$ and $v_{th}$)}
\begin{equation}
    \gamma \sim \rho^s l^{s-1}v_{th}.
\end{equation}

%These time scales are also obtainable from dimensional analysis of the system's parameters.
   {Physically relevant time scales could be identified with integer values of $s\geq0$. $s=0$ gives a time scale $t_c= {l}/v_{th}$ which is proportional to the interaction time $during$ a binary collision. $s=1$ gives the dynamical time scale $t_d$ which is proportional to the mean time $between$ binary collisions.} %Given a particle in a certain region on the ring, the rate of other particles coming into this region to interact with it is ${\rho}v_{th}$.
Similarly, $s=2$ gives the rate of ternary interactions. The probability that two particles existing in an interaction region of length $l$ scales as ${\rho}l$, and the rate of a third particle entering this region to interact with the other two particles is ${\rho}v_{th}$. Thus the ternary interaction rate scales as $\rho^2 l v_{th}$.

For each additional particle colliding, there is one more factor of ${\rho}l$. For small densities where $\rho l \ll 1$, the relevant term for thermalization is the 3-body interaction term since the two body collision is trivial.  In the thermodynamic limit $N\rightarrow \infty$, we get the scaling of the 3-body thermalization rate in Eq. (\ref{eqn: Gamma definition}):
\begin{equation} \label{eqn:Gamma}
    \Gamma = a_n \, \rho^2 l v_{th},
\end{equation}
where $a_n$ is a dimensionless quantity depending only on $n$ in Eq. (\ref{eqn:ipp}). 
%We demonstrate Eq. (\ref{eqn:Gamma}) numerically by two methods: directly by implementing MD simulations, and indirectly by computing the 3-body collision kernel.
%The noisy MD distribution is plotted alongside the Boltzmann prediction discussed in Sec. \ref{sec:MD_BM}.

\begin{figure}[h]
\resizebox{86mm}{!}{\includegraphics{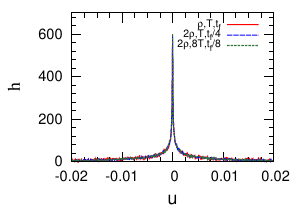}}
\caption{\label{fig:MD_with_den_temp} Three different parameter sets with respective $t_f \propto 1/(\rho^2\,T^{1/3})$  give identical final distributions.}
\end{figure}

From Eq. (\ref{eqn:Gamma}), we expect that the evolution rate is proportional to $\rho^2\,T^{1/3}$ for $n=6$. To test this scaling, we compare a gas with density $\rho$ and temperature $T$ evolving for time $t_f$ with two other cases: $2\rho$, $T$ evolving for $t_f/4$ and $2\rho$, $8T$ evolving for $t_f/8$. Figure \ref{fig:MD_with_den_temp} shows the evolved distribution in the three cases. All three curves are the same within statistical uncertainty, which confirms the predicted scaling. 

Parameters used in the MD simulation are $N=161$ particles for three cases: $R=51200 \,\mu m$, $kT = 100\,meV$, $t_f=12800\,ns$; $R=25600\,\mu m$, $kT = 100\,meV$, $t_f=3200\,ns$; and $R=25600\,\mu m$, $kT = 800\,meV$, $t_f=1600\,ns$. In all cases, $l_0=1\,\mu m$, $\epsilon= 1\,eV$, $m=m_e$, and $n=6$. In all cases, the final time $t_f = 1.6 \times 10^{-4}/\Gamma$ as computed from Eq. (\ref{eqn:Gamma}). 

%\subsection{Random Walk Model}
\subsection{Scaling of the Collision Kernel $K$}
\label{sec:collision kernel}
%\subsection{Scaling of Monte Carlo simulation}
%\label{sec:scaling of MC}
% do not inline anymore man
% Before we present the results of the MC simulations, we provide a scaling argument based on the dimensionless scaled equations of motion of the 3-body collision.
A necessary condition for Eq. (\ref{eqn:CTMC}) to reproduce the MD evolution is that $K\Delta v$ as computed from the MC simulations must have the scaling of $\Gamma$ in Eq. (\ref{eqn:Gamma}). We show that $K$ indeed has the desired scaling under the assumption that $\rho l\ll1$, where $l$ is defined in Eq. (\ref{eqn:length scale}).  If we define $\tilde{x}= x/l$, $u={v}/{v_{th}}$, and $\tilde{t}={t v_{th}}/{l}$, then those equations are

\begin{eqnarray}
\label{eqn:scaled_3b_eom}
        \frac{du_i}{d\tilde{t}} &=& \sum_{j=0, j\neq i}^{2} \frac{n}{(\tilde{x}_i-\tilde{x}_j)^{n+1}}, \\
        \frac{d\tilde{x}_i}{d\tilde{t}}&=&u_i.
\end{eqnarray}

Since the length scale is $l$, we set the length of the region $L$ to scale with $l$. In our calculations, for example, we get converging results for the scaling of $K$ at $L=90l$ and $\rho l= 5 \times 10^{-5}$. The scaled dynamics (Eq. (\ref{eqn:scaled_3b_eom})) of a single trajectory only depends on $n$; that is, scaled initial velocities and positions map to scaled final velocities and positions irrespective of $\rho$ and $T$. If $\rho$ or $T$ changes, the rate $r$ of launching changes according to Eq. (\ref{eqn:rate_launch_zero}), and the average time spent in the observation region scales as $l/v_{th}$. 

%(assuming the initial velocities are scaled in the same fashion)

Not all simulated trajectories result in a non-trivial change in velocities. Only trajectories where the two launched particles coincide for some time in the region result in effective 3-body scattering. Otherwise, the collision is just a sequence of binary collisions that only swap velocities just like in a Newton's cradle. 

% Refree 1 Edit 7
  {This coincidence rate} $\Gamma_{coin}$ is calculated according to a Poisson process with rate $r$ and observation time window $t_{ob}\propto l/v_{th}$.
\begin{eqnarray} \label{eq:rate_coincide}
    \Gamma_{coin}
    &=& \frac{\mathrm{probability\,of\,two\,arrivals}}{t_{ob}} \nonumber \\
    &=& \frac{(rt_{ob})^2\,e^{-rt_{ob}}}{2t_{ob}} \approx \frac{1}{2}r^2t_{ob} \propto \rho^2 v_{th} l
\end{eqnarray}

%\subsubsection{Scaling of K}
 %We check that $K$ has the same scaling as that of $\Gamma$  in Eq. (\ref{eqn:Gamma}) 
 
\begin{figure}[h]
\resizebox{86mm}{!}{\includegraphics{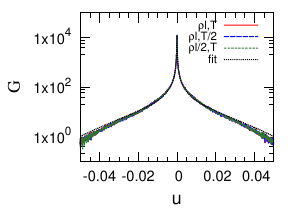}}
\caption{\label{fig:G_tilde_uni}   {The dimensionless scaled collision kernel $G_{0\rightarrow u}$ for different values of $T$ and $\rho l$. The height and shape of all curves are identical which demonstrates the universal scaling. A fitting function $\frac{\alpha}{|u|^{\beta}}e^{-u^2/2\sigma^2}$ is also plotted and explained in Sec.} \ref{sec:Fitting}.}
\end{figure}

 The coincidence rate (rate of effective 3-body collisions) is proportional to the rate of thermalization, $\Gamma$, in Eq. (\ref{eqn:Gamma}) and the kernel $K\Delta v$, which we verify by introducing the dimensionless scaled kernel
 \begin{equation}\label{eqn:normalized kernel}
  G_{u'\rightarrow u} \Delta u= \frac{K_{v'\rightarrow v} \Delta v}{\rho^2 l v_{th}},
 \end{equation} 
 where $u=v/v_{th}$. We show in Fig. \ref{fig:G_tilde_uni} that $G$ is independent of $T$ and $\rho$ given that $\rho l\ll1$, $L/l\gg 1$. The scaled kernel $G_{0\rightarrow u}$ (denoted by $G$ in the y-axis label) is plotted against the scaled velocity $u = v/v_{th}$. The parameters used in the MC simulations are $v'=0$ (initially stationary special particle), $n=6$, $kT = 100\, meV$, $l_0=1\,\mu m$, $\epsilon= 1\,eV$, $m=m_e$, bin width $\Delta v = 0.1v_{th}/2000$, $\rho l = 5\times 10^{-5}$ and $L=90l$, where $l$ is defined in Eq. (\ref{eqn:length scale}). 

% Refree 1 Edit 8

Since $G$ is independent of $\rho$ and $T$, the kernel scales like $K \Delta v \propto \rho^2 l v_{th}$. Although we do not explicitly show the scaling with the other parameters ($m$,$\epsilon$, $l_0$), it is implied in the definition of $l$ and $v_{th}$ as described in the scaling argument. The scaling of the kernel $K$ with most of the system parameters $\{m,\epsilon, l_0, T, \rho\}$ means that effectively we only need to simulate one representative case for different values of the remaining parameters $\{u', n\}$ in order to cover the whole parameter space. We implement this idea in a later section.

% All the velocities are scaled by $v_{th}$. In the graph axis labels $G(u)$ means $G_{u'\rightarrow u}$. 
%We work with the scaled kernel $G$ instead of $K$ to demonstrate the scaling behavior anticipated in the Sec. \ref{sec:scaling of MC}.

%check normalization by v_th here. I should just vtilde right

% \[
% \begin{split}
%     r
%     &=\frac{d\langle\tilde{v}^2\rangle}{dt} \vert_{t=0} \\
%     &= \sum_{i=-B}^{i=B} \sum_{j=-B}^{j=B} G_{j\rightarrow i} D_j(0) \tilde{v}_i^2 \Delta \tilde{v} \\
%     &= \sum_{i=-B}^{i=B} \sum_{j=-B}^{j=B} G_{j\rightarrow i} \delta_{j0} \tilde{v}_i^2 \\ 
%     &= \sum_{i=-B}^{i=B} G_{0\rightarrow i} \tilde{v}_i^2 \\
%     &= \left(\sum_{i=-B}^{i=B}  \tilde{G}_{0\rightarrow i} \tilde{v}_i^2\right) \rho^2 L v_{th} \\
%     &= a_6 \rho^2 L v_{th}
%     \mathrm{,}
% \end{split}
% \]
%
Using the generated values for $G$ in Fig. \ref{fig:G_tilde_uni}, the rate $\Gamma$ in Eq. (\ref{eqn:Gamma}) can be computed using Eq. (\ref{eqn:Dis_CTMC}) and Eq. (\ref{eqn:normalized kernel}) as the rate of change of the variance of $h$:
\begin{eqnarray} \label{eqn: rate initial}
    \Gamma
    &=& \left.\frac{d\langle (u-u')^2\rangle_h}{dt} \right\vert_{t=0}\nonumber \\
    &=& \rho^2 l v_{th} \sum_{j=-\infty}^{j=\infty} G_{u'\rightarrow u_j} (u_j-u')^2 \,\Delta u   \nonumber\\
    &=& a_n \rho^2 l v_{th},
\end{eqnarray}
so the proportionality constant $a_n$ is given by the variance of $G$:
\begin{equation} \label{eqn:prop constant}
    a_n = \langle (u-u')^2\rangle_G.
\end{equation} 

%That explains the $n=2$ entry in Tab.\ref{tab:stiffness}.
 
 % The numerical evidence that the potential for the other powers is diffractive (momenta change non-trivially) might be a suggestion that the 3-body problem is non-integrable.

 %get those parameters for other n's?

%More accurately, we average the fitting function over bins of size $\Delta \tilde{v}$ because $G$ is discrete-sized in the same fashion. That will accurately fit the bins adjacent to 0. (do that)

%At $v \approx 0.04v_{th}$, the kernel $G$ drops by around $4$ orders of magnitude from its maximum value at the bin adjacent to zero. That means that for all practical purposes we can consider the kernel to be zero for $v>0.04v_{th}$. 

% integral bounded around v=0

%
%the stiffer the potential, the less effective a collision is in redistributing the energy. I.e.,

%interesting cases for n=2 but takes too much to simulate, and 1 which is 

\subsection{MD and Boltzmann Evolution Comparison} \label{sec:MD_BM}

\begin{figure}[h] 
\resizebox{86mm}{!}{\includegraphics{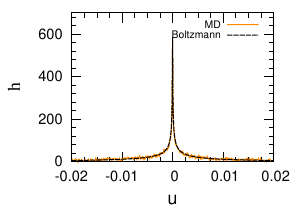}}
\caption{\label{fig:MD_FP} MD and Boltzmann evolved distributions are in agreement. See text for relevant simulation parameters.}
\end{figure}

To show that the kernel, $K$, contains the information of the thermalization dynamics, we use it to evolve $h(v,0)=\delta(v)$ according  Eq. (\ref{eqn:Dis_CTMC}) and compare the evolved distribution to that of the MD simulation. We tested several values of the evolution time $t_f$ and different values for the system parameters. When $N$ is big and $\rho$ is small, the Boltzmann and the MD evolved distributions are in agreement as shown in Fig. \ref{fig:MD_FP}. This demonstrates that the collision kernel, $K$, describes the thermalization process in the low density limit. For the comparison presented here, we use system parameters $N=161$, $R=51200\,\mu m$, $l_0 = 1\, \mu m$,  $n=6$, $\epsilon = 1 \, eV$, $kT = 100 \, meV$, $m=m_e$, and $t_f=12800\,ns = 1.6 \times 10^{-4}/\Gamma $, where $\Gamma$ is computed from Eq. (\ref{eqn:Gamma}) using $\rho=(N-1)/(2\pi R) \approx 0.5\, mm^{-1}$.

%Hence, we use $K$ to study the distribution at longer times.
%Ahmad Haitham, your captions should be very descriptive of the whole figure. They should be standalone things. Try to use Askii or Unicode for special characters in Gnuplot. Should I introuce another t normalized by Gamma. statistical time?

For smaller $N$ or bigger $\rho$, we get a slight disagreement between the two methods. Four-body collisions are significant when $\rho$ becomes large, which is not accounted for in the collision kernel, $K$. Moreover, at fixed $\rho$, the MD simulations require large $N$ for convergence because the perturbation, $h$, has a large effect on the thermal particles for smaller $N$.

%when $N$ is not large enough, the perturbation $h$ is not very small, which through thermalization causes the the temperature of the bath to drop slightly.

%\subsubsection{Boltzmann Evolution for longer times }

Equipped with the kernel, we can evolve the system for longer times using the Boltzmann Eq. (\ref{eqn:Dis_CTMC}). First, we calculated the kernel for a slow moving special particle $u'=-0.25$ and found the kernel $G$ is approximately translation invariant as later demonstrated in Fig. \ref{fig:Dif_vel}. That is,
\begin{equation} \label{eqn:Gappro}
    G_{0\rightarrow u} \approx G_{u'\rightarrow u'+u}.
\end{equation}
for $|u'| \ll 1$. Therefore, the only information needed to propagate small velocities to good accuracy is $G_{0\rightarrow u} $. In this limit, the change in $h$ during successive time steps in Eq. (\ref{eqn:CTMC}) becomes a repeated convolution integral. 

\begin{figure}[h]
\resizebox{86mm}{!}{\includegraphics{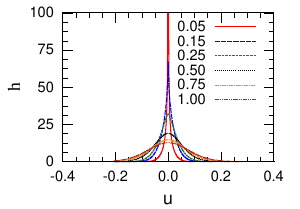}}
\caption{\label{fig:approach_all} Evolution of the perturbation, $h(u,t)$, at different times and approach toward equilibrium. Gas parameters are $N=161$, $n=6$, $R = 51200 \, \mu m$, $\epsilon = 1 \, eV$ and $kT = 100 \, meV$, $m$ is the mass of the electron. Figure legends represent the time for the distribution in fraction of $0.01/\Gamma$, where $\Gamma$ is computed from Eq. (\ref{eqn:Gamma}). There are two stages for the peak of the perturbation: non-Gaussian cusp and smooth Gaussian.}
%In this paper, we explain this evolution of $h$, calculate the rate of thermalization $\Gamma$, and study its scaling with the system's parameters.
\end{figure}

Figure \ref{fig:approach_all} shows the evolution of the delta perturbation over a time scale $t_f= 0.01/\Gamma$ as computed from the Boltzmann Eq. (\ref{eqn:CTMC}). The system's parameters are $N=161$, $R=51200\,\mu m$,  $n=6$, $\epsilon = 1 \, eV$, $kT = 100 \, meV$, $m=m_e$, where $\Gamma$ is computed from Eq. (\ref{eqn:Gamma}) and $\rho=(N-1)/(2\pi R)$. At early times $t_f\Gamma < 1 \times 10^{-3}$, $h$ has a cusp maximum which resembles that of $G$. That is because the change in $h$ is approximately proportional to $G$ as in Eq. (\ref{eqn:CTMC_localized}) when $G$ is highly localized. At later times $t_f\Gamma \sim 1 \times 10^{-2}$, the cusp flattens out as repeated convolutions relax the population to a Gaussian distribution, which subsequently spreads at a steady rate during the range of time considered. The relaxation to a Gaussian is a consequence of the repeated convolution and the Central-Limit theorem\cite{klenke2013probability}.

%There are two distinct stages for the shape of $h$: an early stage with cusp peak that is explainable in terms of the initial delta perturbation and the collision kernel (Sec. \ref{sec:collision kernel}), and a later stage with Gaussian character that is explained by the Central-Limit theorem \cite{klenke2013probability}. This simulation is done through 3-body collisions as explained in Sec. \ref{sec:MD_BM}.

\begin{figure}[h]
\resizebox{86mm}{!}{\includegraphics{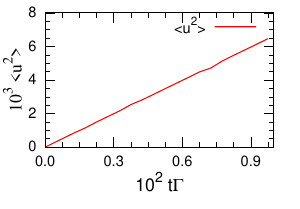}}
\caption{\label{fig:energy} The variance $ \langle u^2 \rangle_h$ of the perturbation growing linearly with time. The evolution of $h$ itself is plotted in Fig. \ref{fig:approach_all}.}
\end{figure}

The spread of the perturbation in Fig. \ref{fig:approach_all} indicates that its variance (which is proportional to the energy) is growing with time which is shown in Fig. \ref{fig:energy}.   {The linear evolution of the perturbation's energy in the time range considered indicates the thermalization rate defined through Eq.} (\ref{eqn: Gamma definition})   {is approximately constant with time and does not depend strongly on the shape of perturbation}. It also suggests modelling the thermalization process as a random walk of the velocity $u'$ of the special particle. The standard deviation of $h$ is proportional to $\sqrt{t}$, just like the standard deviation of displacement in a random walk is proportional to the square root of the number of steps taken.

Each 3-body collision with the thermal particles gives the special velocity a random kick leading to a random walk in velocity space. Using the order of magnitude of $a_n$ in Tab. \ref{tab:stiffness}, it takes $\sim 1/a_n \sim 10^4$ kicks to thermalize, i.e., the variance $ \langle u^2 \rangle$ approaching $1$. The kick magnitude is $\sim \sqrt{a_n} \sim 0.01$, which is of the order of magnitude of the strongest kick in a 3-body collision (as explained in Sec. \ref{sec:Fitting}).

\subsection{G for a moving special particle $u'\neq 0$}
\begin{figure}[h]
\resizebox{86mm}{!}{\includegraphics{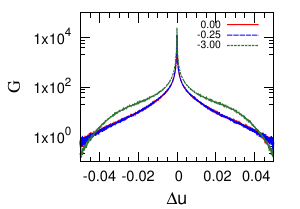}}
\caption{\label{fig:Dif_vel} Kernel $G$ for different special velocities. A perturbation at the tail scatters faster than a perturbation at the center. The $u'=-3.0$ curve is slightly skewed to the right as shown in the first moment $ \langle u \rangle$ in Fig. \ref{fig:v_v2_drift}. The $x$-axis here is $\Delta u = u - u'$.}
\end{figure}

%Now that we have analyzed the scaling of the thermalization process in two ways
In this section, we demonstrate how other velocity perturbations $u'\neq 0$ (moving special particles) scatter within the bath due to 3-body collisions. Since the kernel is proportional to the transition rates, Fig. \ref{fig:Dif_vel} shows that a perturbation near the tail of the Maxwell-Boltzmann distribution at $u'=-3.0$ scatters to neighboring velocities more rapidly than from $u'=0$. On the other hand, scattering from small velocities such as $u'=-0.25$ is almost identical to $u'=0$.

%The $-3v_{th}$ curve is shifted $3$ units to the right to compare with the $0$ curve.

\begin{figure}[h] 
\resizebox{86mm}{!}{\includegraphics{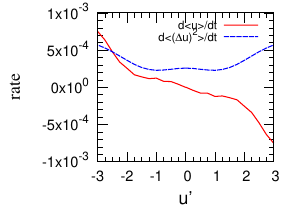}}
\caption{ \label{fig:v_v2_drift} Initial rate of change (scaled by $\rho^2 l v_{th}$) of the first moment $ \langle u \rangle$ and the variance $ \langle (u-u')^2 \rangle$ for perturbations localized at different velocities. The calculations were done for negative $u'$, but were reflected to extend over positive $u'$ for clarity. The small fluctuations in the curves are due to statistical noise.}
\end{figure}

For several values of  $u'$, the initial rate of change of the mean scaled velocity $d\langle u \rangle/dt$ and the variance $d\langle (u-u')^2 \rangle/dt$ are calculated as in Eq. (\ref{eqn: rate initial}) and are shown in Fig. \ref{fig:v_v2_drift}. In particular, negative velocity perturbations have a positive initial rate of change of the average, which is a drag effect that slows down the special velocity. Also, we note that the initial rate of change of the variance only changes by a factor of $\sim 2$. This indicates that the definition of $\Gamma$ in Eq. (\ref{eqn: Gamma definition}) leads to a reasonable estimate of the time required for thermalization (i.e., $\Gamma$ does not strongly depend on $u'$).

From the $\langle (\Delta u)^2 \rangle$ curve, bigger velocity perturbations transition faster to neighboring velocities. This is an indication of detailed balance \cite{gorban2014detailed} which we numerically checked for pairs of velocity states. For example, the bins at $u=-3.00$ and $u=-2.96$ exchange populations at rates compatible with the steady state Maxwell-Boltzmann distribution. That is, for $n=6$ we numerically found that

\begin{equation} \label{detailed balance}
    \frac{{G}_{-3.00\rightarrow-2.96}}{{G}_{-2.96\rightarrow-3.00}} = 1.12 \pm 0.04,
\end{equation} while the ratio between the Maxwell-Boltzmann populations at the corresponding bins is $ {e^{-{2.96^2}/{2}}}/{e^{-{3.00^2}/{2}}} = 1.13.$

% seriously let's use v tilde again or u for that matter. 

%We, however, do not study the evolution near the thermalization time $t_f\Gamma \sim 1$ at which the variance saturates at 1. To study thermalization very close to equilibrium, Eq. (\ref{eqn:CTMC}) requires the full information of the kernel. 

\subsection{Dependence of $G$ on the Potential Power $n$} \label{sec:Fitting}
\begin{figure}[h]
\resizebox{86mm}{!}{\includegraphics{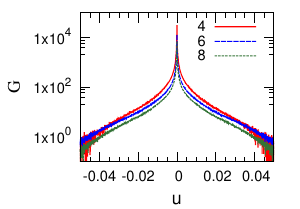}}
% Here is how to import EPS art
\caption{\label{fig:G_tilde_different_n} The collision kernel $G_{0\to u}$ for different potential power $n$ values. $G_{0\to u}$ is generally bigger for smaller $n$ for the range of $u$ shown.}
\end{figure}

Figure \ref{fig:G_tilde_different_n} shows how the kernel compares for different powers $n$ in the inverse power potential, Eq. (\ref{eqn:ipp}). The scaled kernel is generally smaller for bigger $n$, indicating that the special particle scatters more slowly when the potential is steeper. Particularly, in the limit $n \rightarrow$ $\infty$, the potential in Eq. (\ref{eqn:ipp}) approaches hard walls at $|d|=l_0$, and the binary collision is an instantaneous velocity swap of two particles of size $2l_0$. In such limit, a ternary coincidence  (Eq. (\ref{eq:rate_coincide})) necessary for thermalization is impossible. This is reflected in the decrease of $a_n$ with increasing $n$, where the values of $a_n$ computed using Eq. (\ref{eqn:prop constant}) are shown in Tab. \ref{tab:stiffness}.

%\subsubsection{Fitting the Collision Kernel} \label{sec:Fitting}
To understand why the kernel has its shape and why different values of $n$ produce different shapes, we fitted the kernel guided by the details of the MC simulation. In Fig. \ref{fig:Dif_vel}, the kernel $G_{-3.0\rightarrow -3.0+\Delta u}$ near $\Delta u = 0$ decreases rapidly with increasing $|\Delta u|$, and decreases more rapidly for $|\Delta u|>\sim 0.03$. To capture both of these features, $G$ can be fitted to a Gaussian modulated power law: 
\begin{equation} \label{eqn:fit}
  g(u)=\frac{\alpha}{|u|^{\beta}}e^{-u^2/2\sigma^2}  
\end{equation}
 as shown in Fig. \ref{fig:G_tilde_uni} for $n=6$. The parameters $\alpha$, $\beta$, and $\sigma$ are shown for several values of $n$ in Tab.\ref{tab:stiffness}. The $\alpha$ parameter is proportional to the over all scattering rate, and the power $\beta$ measures how fast the kernel drops with increasing $|u|$ near $u=0$ as in Fig. \ref{fig:G_tilde_uni}. We observe that both $\alpha$ and $\beta$ decrease as $n$ increases (i.e., when the potential is becoming steeper and approaching hard walls).
  
\begin{figure}[h]
\resizebox{86mm}{!}{\includegraphics{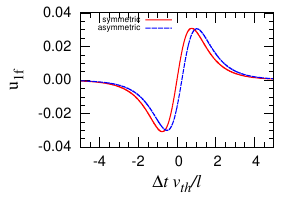}}
\caption{\label{fig:scattering} Scaled kick $u_{1f}$ as a function of the scaled delay ($\Delta t v_{th}/l$) between launches. The symmetric case is for $u_{2i}=1$ and $u_{3i}=-1$, while the asymmetric case is for $u_{2i}=1.1$ and $u_{3i}=-0.9$. The curve for the asymmetric case was shifted horizontally for clarity. Both curves have long tails and peak at $u_{1f,max}$ $\sim 0.03$.}
\end{figure}

The choice of such fitting function can be explained by inspecting mono-energetic collisions in which the incoming energies are fixed as opposed to having a continuous `thermal' distribution. Figure \ref{fig:scattering} shows the effect on the special particle starting from zero velocity in two launching cases: a symmetric case $v_2=v_{th}$, $v_3=-v_{th}$, and an asymmetric case $v_2=1.1v_{th}$, $v_3=-0.9v_{th}$. The final kick the special particle receives, $u_{1f}$, is plotted against the delay $\Delta t$ between launching particles 2 and 3. Both curves look similar with long tails for $|\Delta t| \gtrsim 3l/v_{th}$. The long tails happen because a large delay results in a tiny momentum transfer to the special particle. They are responsible for the fast drop of the kernel $G_{0\to u}$ around $ u=0$, which is captured in the power-law term of the fitting function $g$. Moreover, both curves peak at a maximum momentum transfer $u_{1f,max}$ $\sim 0.03$. The Gaussian modulation term in Eq. (\ref{eqn:fit}) is a way to average over the distribution of the launched particles from the thermal environment. Its width $\sigma$ is not of order $1$ but rather reflects the value of the fractional momentum transfer $u_{1f,max}$. In particular, $\sigma$ correlates with $|u_{1f,max}|$; they both peak at $n = 6$ and drop monotonically away from $n = 6$ as shown in Tab. \ref{tab:stiffness}.

%The kernel in this case is denoted by $G_e$ and computed from an algorithm similar to that in Appendix \ref{Appendix:MC Algorithm}.  $G_e$  can fitted to a power law with a cut off at certain maximum velocity $u_{1f,max}$. $u_{1f,max}$ is the biggest fractional momentum transfer during a mono-energetic collision and corresponds to a certain delay between the launching of particles 2 and 3.

\begin{table}[h]
\caption{Rate constants $a_n$ and fitting parameters for different potential powers $n$. The uncertainty in $\alpha$, $\beta$, and $\sigma$ is $\pm0.001$.}
\label{tab:stiffness}
\begin{ruledtabular}
\begin{tabular}{c c c c c c}
$n$ & $a_n$ & $\alpha$ & $\beta$ & $\sigma$ & $|u_{1f,\text{max}}|$ \\
\hline
2   & 0                      & 0.000 &         &         & $1 \times 10^{-7}$ \\
4   & $3.49 \times 10^{-4}$  & 0.237 & 1.182   & 0.022   & $2.69 \times 10^{-2}$ \\
6   & $2.58 \times 10^{-4}$  & 0.182 & 1.117   & 0.027   & $3.06 \times 10^{-2}$ \\
8   & $1.62 \times 10^{-4}$  & 0.148 & 1.076   & 0.025   & $2.76 \times 10^{-2}$ \\
10  & $1.02 \times 10^{-4}$  & 0.129 & 1.046   & 0.021   & $2.37 \times 10^{-2}$ \\ 
12  & $6.55 \times 10^{-5}$  & 0.115 & 1.023   & 0.018   & $2.04 \times 10^{-2}$ \\
\end{tabular}
\end{ruledtabular}
\end{table}

For $n=2$, $|u_{1f,max}|$ was found to be $\approx 1\times 10^{-7}$ which is not different from $0$ within errors resulting from the numerical solution of the equations of motion, Eq. (\ref{eqn:scaled_3b_eom}). In fact, for the inverse square power potential it was shown that the equations of motion are integrable and the potential is, surprisingly, isospectral (momenta only trivially swap) \cite{calogero1974exact}. That means that a system with such pairwise interaction can only thermalize through four or higher body collisions.

%The resolution is that even if the scattering rate in infinite, the rate of change of perturbation's velocity $ \langle u \rangle_g$ - the first moment of $g$ - is bounded as demonstrated in Fig. \ref{fig:v_v2_drift}.

\subsection{Broken Scaling in Lennard-Jones potential} \label{sec: LJ}
In constrast with the inverse power law potential in Eq. (\ref{eqn:ipp}), the Lennard-Jones potential is 

% { [ ( ) ] }
\begin{equation}
    \label{eqn:LJ}
    U_{LJ}(d)=\epsilon \left[\left(\frac{l_0}{d}\right)^{12}-\left(\frac{l_0}{d}\right)^6\right].
\end{equation} 
This potential is different from the inverse power potential in important ways. First, it is attractive at long distance. Moreover, there is possibility of 3-body recombination.
%find paper and compare quantum 3d, 1d 3b recombination with 1d recombination rates.
Most importantly, the scaling of the inverse power potential is lost in the Lennard-Jones potential. That is, we cannot fix the potential energy scale to $kT$ as we did in Eq. (\ref{eqn:inverse power potential-kT scaled}).

\begin{figure}[h]
\resizebox{86mm}{!}{\includegraphics{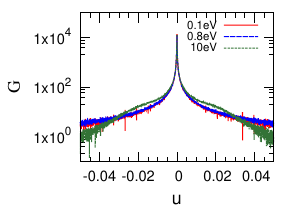}}
\caption{\label{fig:LJ} Scaled collision kernel for different $kT$ with $\epsilon = 1eV$ and other parameters from Sec. \ref{sec:collision kernel}. The small temperature curves overlap, but the high temperature curve deviates indicating the lack of scaling symmetry.}
\end{figure}

Figure \ref{fig:LJ} shows $G_{0 \to u}$ for the Lennard-Jones potential for three different temperatures $kT= 0.1, 0.8, $ and $10 \, eV$. In all cases, the potential energy scale $\epsilon = 1\, eV$. When the kinetic energy ($kT$) is small compared to the potential energy, the scaled $G_{0 \to u}$ is almost the same for different temperatures as seen in the $0.1$ and $0.8 \, eV$ curves. When the kinetic energy is larger, $G_{0 \to u}$ has a significantly different shape as seen in the $10\,eV$ curve.
This shows the universal scaling (compare to Fig. \ref{fig:G_tilde_uni}) remains approximate at low $T$ but is lost at high $T$. The simulation parameters are the same as those used for Fig. \ref{fig:G_tilde_uni}. The $G$ kernel here is still scaled by $l$ defined through Eq. (\ref{eqn:length scale}) with $n=6$. For $kT= 100 \, meV$, the dimensionless thermalization rate $a_n$ is found to be $3.00 \times 10^{-4}$, which is bigger than that of the power law potential with $n=6$ by $16 \%$. Thus, the Lennard-Jones potential gives qualitatively similar thermalization rates to the inverse power potential even with the differences noted above.

It is interesting to compare the rate of 3-body recombination to the rate of thermalization for 1D and 3D gases. In the present 1D gas, both processes involve three particles, and their rates scale like $\rho^2$. 
In dilute 3D gases, thermalization is a 2-body process (with rate $\propto \rho$) and happens at a much faster rate than 3-body recombination. Moreover, for fixed $\rho$ in our 1D gas, we find that the ratio of 3-body recombination rate to thermalization rate drops rapidly with increasing $T$. This could be explained by noting that at low $T$, a pair of particles approaching each other from far distance have small positive relative energy. If a third particle interact with the pair, there is a high chance that the energy of the pair transfers to the third particle, leaving the pair in a bound state with negative relative energy. At high $T$, the chances of the third particle taking away enough energy to switch the sign of the energy of the pair is small.

\section{Conclusion}

We have studied the rates of classical three-body thermalization in dilute one-dimensional gases with inverse power law interaction with $n>2$ for various system parameters. Through MD simulations of the $N$-body gas and MC simulations of the 3-body scattering kernel, we showed that the gas relaxes with a rate proportional to $\rho^{2} \sqrt{\frac{kT}{m}}l_0(\frac{\epsilon}{kT})^{\frac{1}{n}}$. The scaling of the thermalization rate in terms of $T$ is exact for the inverse power potential, but only approximate at low $T$ for the Lennard-Jones potential.

Classical 3-body thermalization in 1D is significantly slow (compared to higher dimensional gases) not only due to the $\rho^2$ scaling, but also the smallness of the proportionality constant, which comes from the weak redistribution of energy in each collision. The implication is that dilute 1D gases preserve their out-of-equilibrium states for a relatively long time. For example, a 1D Nitrogen atom gas with $\rho = \,1$ atom$/10 \,\mu m$ at $T=10 \,K$ interacting under the Nitrogen-Nitrogen Lennard-Jones potential takes around $10 s$ to thermalize according to Eq. (\ref{eqn: rate initial}) and parameters found in \cite{LJ}.

By rewriting the equations of motion in a dimensionless way, we provided arguments for the scaling of both MD and MC simulations, which we verified numerically. Moreover, we have shown that the Boltzmann equation using a three-body collision kernel is sufficient to reproduce the evolution of low density $N$-body gas calculable from MD simulations. Additionally, the collision kernel behaves like a power law for small momentum transfer. For bigger momentum transfer, it is modulated by a Gaussian with width of order of the maximum momentum transfer during a mono-energetic collision.
Finally, the collision kernel provided us with an understanding of how the overall rates and the statistical details of three-body scattering depend on the potential power $n$.
\\

Data for the figures used in this publication are available
from the Purdue University Research Repository \cite{data}.

\begin{acknowledgments}

This work was supported by the US Department of Energy, Office of Science, Basic Energy Sciences, under Award No. DE-SC0012193.
\end{acknowledgments}

\appendix

\section{\label{sec:MD simulation}Molecular Dynamics Simulation} 

In this appendix, we provide details and analysis of the MD simulations. To simulate the gas of $N$ particles in Fig. \ref{fig:ring}, the initial velocities are chosen as per Eq. (\ref{eqn:initial distribution}), whereas the initial locations are chosen randomly according to a relative potential energy Boltzmann factor. We evolve the system from time $t_i=0$ to time $t_f$ according to Eq. (\ref{eqn:dynamical equations ring}) using Runge–Kutta methods with adaptive time step \cite{press2007numerical}. At $t_f$, we subtract the background thermal distribution as per Eq. (\ref{eqn:initial distribution}) to single out the perturbation $h(v,t_f)$.  {  {We average over many trajectories with different initial conditions and average over a window of time around $t_f$ to reduce the statistical noise.}}

% Refree 2 Second Question
 {  {The number of terms in the force calculation in Eq.}} (\ref{eqn:dynamical equations ring})  {  {scales like $N^2$ for a single time step, which makes it difficult to simulate more than 30 particles. To decrease the computational time, we utilize that the inverse power potential in Eq.}} (\ref{eqn:ipp})  {  {is relatively `short-range' for $n \geq  2$ and small density $\rho$. In this regime, particles are only interacting significantly with their close neighbors, while the interaction with further particles can be ignored. This allows us to do nearest neighbors calculations for the force, which results in a time complexity that scales like $N$. This approximation is similar to truncating the potential at certain particle separation as usually used in the MD simulation for the Lennard-Jones potentials}}\cite{giordano2012computational}. The neighbors are selected by ordering the particles according to their initial locations. For 3 nearest neighbors, for example, particle number 5 experiences a force from particles numbered 2-4 and 6-8. Since the potential is infinitely repulsive at short distance, the particles cannot pass through each other and the particle order is fixed. We checked that the results converged for 3 nearest neighbors which was used for all MD simulations in this paper. Using this method, we could simulate more than 160 particles.

To ensure that 3-body collisions are the largest effect, we choose a small density $\rho = (N-1)/2\pi R$. We verify the $\rho^2$ scaling by simulating a density $\rho$ for time $t_f$ and comparing the final $h$ to that of another simulation with $\rho/2$ (by doubling $R$ for a given $N$) for time $4t_f$. The results converge for densities near $(\rho l \sim 1\times 10^{-3})$. From analysis of the Runge-Kutta with adaptive time step algorithm, the related best-case time complexity scales like $1/\rho$.

Once we fix the convergence density, we double $N$ and $R$ in steps to obtain thermodynamic convergence. The results converge for particle number near $N=160$. Once we fix the convergence particle number, we run the simulation for a longer time to get appreciable evolution of $h$. All these considerations combined render the MD simulations computationally expensive, and we chose to limit the simulation time to $t_f$ of the order $1\times 10^{-4}/\Gamma$, where $\Gamma$ is from Eq. (\ref{eqn:Gamma}).

\section{\label{sec:Boltzmann equation} Boltzmann Equation } \label{app:Boltzmann Equation}
%Refer to paper\\
%symmetries of the 3 body collisions\\
%extreme cases of launch velocity and times\\
%which particle is special\\
%guts and details of thermlization as in paper GaAs

In the kinetic theory of gases, the state of a gas in 1D is described by the aggregate one-particle distribution in coordinate-velocity space $f(x,v,t) \,dx dv$. The evolution of such distribution is described by the Boltzmann equation \cite{Zwanzig_2009}:

\begin{equation} \label{eqn:Boltz}
    \frac{\partial f}{\partial t}+v\frac{\partial f}{\partial x} + \frac{f}{m}\frac{\partial f}{\partial v}  = \left(\frac{\partial f}{\partial t}\right)_{c},
\end{equation} where $f$ is the external force and $m$ is the mass of the particle. For an isolated system with uniform density $\rho$, which we assume in our work, the velocity and space derivative terms drop out. In this case, the only way to change $f$ is through inter-particle interactions dictated by the collision term on the right hand side. (Henceforward, we use $f$ to represent the distribution in velocity only. The one-particle velocity-space distribution is $\rho f$, where $\rho$ is the number density.)   The collision term is modelled and computed in the MC simulation in Appendix \ref{Appendix:MC Algorithm}.

\section{\label{Appendix:MC Algorithm} Monte Carlo Simulation}

In this Appendix, we provide the relevant distributions and steps of the Monte Carlo Algorithm for generating the discrete version of $K_{v'\rightarrow v}$. First, we treat the case of an initially stationary special particle, $v'= 0$. The one-particle phase space thermal distribution is given by $\rho f_0$ where $f_0$ is the equilibrium velocity distribution in Eq. (\ref{eqn:MB distribution}). At $x=L/2$, particles are entering the observation region with negative velocity. The rate of entry is equal to the probability current (flux) $\rho f_0 |v|$ integrated from $v=-\infty$ to $0$.  We get the same rate from the left. So overall we get a rate of  
\begin{equation} \label{eqn:rate_launch_zero}
     r= \int_{-\infty}^{+\infty} \rho f_0(v)|v| \,dv= \rho v_{th} \sqrt{\frac{2}{\pi}},
\end{equation}
 %\begin{equation} \label{eq:rate}
     %r=\rho \sigma \sqrt{\frac{2}{\pi}}.
 %\end{equation}
%The time that a particle launched with speed $v$ spends in the region after launch is $\frac{L}{v}$; therefore, the probability to find a particle of velocity $v$ in the region is proportional to $\frac{L}{v}f(x,v)v$ which gives a Gaussian distribution of $v$.
and a velocity distribution of the launched particles
\begin{equation} \label{eqn:launch_prop_zero}
 P_{launch}(v) = \rho f_0(v) |v|/r,    
\end{equation}which is the probability that $v$ lies between $v-dv/2$ and $v+dv/2$ per $dv$ (normalized so the integral over all $v$ is 1).
%The average time spent in the region is $<L/v> = {L}/{v_{th}}\sqrt{{\pi}/{2}}$. The average number of particles in the region is, therefore, $r<{L}/{v}> = L\rho$ like expected.
%\rho\sigma{\sqrt{\frac{2}{\pi}}} \frac{l}{\sigma}\sqrt{\frac{\pi}{2}}

%To numerically check that we get back the right density $\rho$ and  the Gaussian distribution in the region, we implement the following test. 1) Generate a sequence of times $t_1, t_2, ..., t_N$ where the separation $\Delta t_j = t_{j+1} - t_j$ follows a Poisson distribution with an average $\delta t_{avg} = \frac{1}{r}$. 2) For each time generate a velocity $v$ according to $P(v)$. Add up $\frac{l}{v}$ and divide by $t_N$. As $N \rightarrow \infty $that should approach the expected number of particles $\rho l$. 3) In a suitable histogram, bin down the $\frac{l}{v}$. That should approach a Gaussian distribution. (Results, what happens when v'!=0)
To treat the case $v'\neq 0$, we go to the reference frame moving with $v'$. The distribution in that frame is $\rho f_0(v+v')$;  therefore, the launching rate is generally
\begin{eqnarray} \label{eqn:rate_launch_nonzero}
       r
     &=& \int_{-\infty}^{\infty} \rho f_0(v+v')|v| \,dv \nonumber \\ 
     &=& \frac{\rho v_{th}}{\sqrt{2\pi}} \int_{-\infty}^{\infty} e^{-(u+u')^2}|u| \,du \nonumber \\
     &=& \frac{\rho v_{th}}{\sqrt{2\pi}} \left[2 e^{\frac{-u'^2}{2}} + \sqrt{2\pi} u' \, erf\left(\frac{u'}{\sqrt{2}}\right)\right],  
\end{eqnarray}
where $erf$ is an error function. The general launch distribution for velocity is
\begin{equation} \label{eqn:launch_prop}
 P_{launch}(v) = \rho f_0(v+v') |v|/r.    
\end{equation} The rate $r$ determines the probability distribution for a delay time between the launching of particles 2 and 3:
\begin{equation} \label{eqn:PDF_delay}
    P_{delay}(\Delta t) = r e^{-r\Delta t}.
\end{equation}

Using these distributions, we implement the MC simulation as follows: 1) At time $t=0$, particle 1 is initialized at $x=0$ with $v_1= 0$. 2) Particle 2 is initialized with a random velocity $v_2$ picked from the distribution $P_{launch}(v)$ in Eq. (\ref{eqn:launch_prop}) at $x=\pm L/2$ depending on the sign of $v_2$. 3) Particle 3 is initialized with a random velocity $v_3$ picked from the distribution $P_{launch}(v)$ at $x=\pm L/2$ depending on the sign of $v_3$, with a random time delay $\Delta t$ chosen according to $P_{delay}(\Delta t)$ in Eq. (\ref{eqn:PDF_delay}). 
% should you talk about the propagating backward with the interaction-less Hamiltonian,?
4) The 3 initial velocities $v_1$, $v_2$, and $v_3$ are added to the appropriate bins (histogram) of the discretized approximation of $K$, Eq. (\ref{eqn:Dis_CTMC}), with a value of $-1$ because these velocities are destroyed through the collision. %with appropriate bin width $\Delta v = v_{th}/N_{bins}$
5) Using Runge-Kutta methods, the  particles are propagated until they collide and separate appreciably. %Choose $\delta t$ to be proportional to the smallest interaction time scale $L/max(v_1,v_2)$ and the final time $t_f= max(\frac{l}{v_1},\frac{l}{v_2}+\Delta t$ to be the time to clear the scene.
6) The final velocities $w_1$, $w_2$,  and $w_3$ are added to the histogram with a $+1$ because these velocities are created.
%(why does the subtraction method, (or selecting the special particle) works? refer to Boltzmann equation). %mention 
7) Steps 1-6 are repeated for a number of trajectories $N_{traj}$ until the statistical noise decreases to a sufficient level. %We accept the noise level in histogram if the noise to signal ratio in the bin at $v=\Delta v/2$ is of order $0.1\%$. 
8) The histogram is divided
%(what do you mean?, write a formula like the scattering cross section)
by $N_{traj}t_{avg}\Delta v$ where $t_{avg}$ is the inverse of the rate $r$. This discretized approximation converges to $K_{v'\rightarrow v'+v}$ in the limit $N_{traj}\to\infty$ and $\Delta v\to 0$. %The statistical noise goes as $1/\sqrt{N_{traj}}$.
%should you talk about normalizing the x-axis of the histogram? you can use \tilde{v} if you know how to insert it in Gnuplot

This prescription is a Monte Carlo evaluation of the scattering kernal where our delta-perturbative model of the collision term can be read from Eq. (\ref{eqn:CTMC}) as 
\begin{equation}
    \left(\frac{\partial h}{\partial t}\right)_{c} = \int_{-\infty}^{\infty} K_{v'\rightarrow v} h(v',t) \,dv',
\end{equation}
where $K$ is computed from the MC simulation. For $v' = 0$, 
\begin{equation} \label{eqn:Boltzdeltas}
    \begin{aligned}
        K_{0\rightarrow v} = &\int\int\int dv_2dv_3d\Delta t  \frac{\rho f_0(v_2)|v_2|}{r}  \frac{\rho f_0(v_3)|v_3|}{r} \\
        & \times re^{-r\Delta t} \times r[-\delta(v)-\delta(v_2-v)-\delta(v_3-v) \\
        & +\delta(w_1-v)+\delta(w_2-v)+\delta(w_3-v)].
    \end{aligned}
\end{equation}
where $w_1, w_2, w_3$ are the outgoing velocities of the collision and are functions of the incoming velocities $v_2$, $v_3$ and the time delay $\Delta t$. $r$ is the rate of launching in Eq. (\ref{eqn:rate_launch_zero}), and $f_0$ is the Maxwell distribution in Eq. (\ref{eqn:MB distribution}). The first three terms in the integrand are the normalized distributions (integral over the respective domain being 1) of $v_2, v_3$, and $\Delta t$. The last term ($r \times$the square bracket) is the rate of destruction subtracted from the rate of creation of velocity $v$. For special velocity $v'\neq 0$, we get a similar integral expression but with the velocity arguments shifted as in Eq. (\ref{eqn:launch_prop}) and $r$ defined in Eq. (\ref{eqn:rate_launch_nonzero}).

Our collision term is comparable to that in Ref. \cite{uhlenbeck1963lectures} which is derived for 2-body collisions in 3D using the differential cross-section. It is also comparable to that in Ref. \cite{ma1983one}, which handles 3-body collisions in 1D but assumes constant transition rates for all $v_1$, $v_2$, $v_3$ $\rightarrow$ $w_1, w_2, w_3$ interactions compatible with energy and momentum conservation. The collision term in Ref. \cite{ma1983one} yields an analytically solvable Boltzmann Equation, but is not derivable from an inter-particle interaction.
%https://en.wikipedia.org/wiki/Stochastic_matrix
%\bibliographystyle{unsrt} % Use for unsorted references
%\bibliographystyle{plainnat} % use this to have URLs listed in References
\bibliography{main}% Produces the bibliography via BibTeX.
%\cleardoublepage
%\bibliography{bibliography} % bath to your References.bib file
%\nocite{\times }
\end{document}